\documentstyle[12pt]{article}
\def\be{\begin{equation}}
\def\ee{\end{equation}}
\def\bea{\begin{eqnarray}}
\def\eea{\end{eqnarray}}

\begin{document}

\begin{center}
{\Large \bf The infinite turn and speculative explanations in cosmology}
\end{center}
\vspace{.2in}
\normalsize
\begin{center}
{\sc Reza Tavakol$^{1}$ and Fabio Gironi$^{2}$} \\
\end{center}
\normalsize
\begin{center}
{\em $^{1}$ School of Physics and Astronomy, \\
Queen Mary University of London,\\
Mile End Road, \\ London. E1 4NS. UK \\
\vspace{.1in}
$^{2}$ School of Philosophy, \\
University College Dublin, \\
Newman Building, \\
Belfield, Dublin 4, Ireland}

\end{center}
\vspace{1cm}
\date{\today}
\begin{abstract}

Infinity, in various guises, has been invoked recently in order 
to `explain' a number of important questions regarding observable phenomena in science,
and in particular in cosmology.
Such explanations are by their nature speculative. Here we introduce the
notions of {\em relative infinity}, {\em closure}, and {\em economy of explanation} and ask: to what 
extent explanations involving relative or real constructed infinities can be treated as reasonable?
\end{abstract}
\section{Background}
 
An important ingredient of many scientific explanations of observable phenomena, including those pertaining to the Universe itself as a whole, have been the so-called unobservable(s), i.e. variables or concepts entering a theoretical explanation without a phenomenologically manifest counterpart, or theoretical entities lying beyond the limits of (current) observability, about which the theory in question is taken to make true claims and which play an indispensable causal-explanatory role.
These have been the subject of a great deal of debate, and in particular whether theories making explanatory use of such unobservable theoretical entities (UTEs) can be treated as complete, or as literally true (as opposed to merely instrumentally efficacious). Despite these debates, it is fair to say that most scientists (as well as the majority of philosophers of science, adhering to scientific realism) have come to terms with the theoretical indispensability and existence of such entities, provided the theory under consideration is {\em currently complete}, i.e. capable of unambiguously accounting for the current observations in its domain of applicability, as well as making novel testable predictions.\footnote{It is important to note that, as opposed to {\em current completeness}, the notion of {\em all time completeness} of theories is problematic in principle, given that all theories need to stand to be corrected by future observations and, in this sense, must remain open.}
 
Such unobservable theoretical entities (UTEs) can be broadly divided into four categories. 
The first consists of those that, though unobservable at a certain epoch, later become observable. 
Subatomic particles\footnote{It is important to differentiate between those subatomic particles that are 
directly visible, such as electrons, and those that are only "observable" in an inferential way, as is the case with most such particles - such as Higgs. Such differentiation does not construe observables and unobservables as ontologically distinct (thus opening the door to instrumentalist skepticism);
the different terms merely reflecting a methodological difference in our access to them (non-inferential or inferential).} are examples of this type of UTEs.

The second  category consists of concepts which are by definition/construction
unobservable and yet lead to observable consequences, such as the wave function
in quantum mechanics, which even though it itself does not have a direct observable
counterpart its square defines probabilities which are measurable.

The third category consists of speculative UTEs that are 
invoked as metaphysical extensions of a successful and predictive theory in order to
give alternative interpretations of how the theory can account for certain observations.
A prime example of this is the so called `many worlds' interpretation of Quantum
Mechanics due to Everett \cite{Everett} which, while denying the reality of 
the collapse of the wave function assumed in usual interpretations of 
quantum mechanics, accords reality to the wave function as well as 
all possible histories, postulating that each represents a different 
physically real and non-interacting 'world'. 

Finally, the fourth category consists of speculative UTEs that are invoked in order to explain an observed aspect of the Universe within an {\em incomplete} theory. The infinite (or very large) number of universes assumed by some current versions of the `Multiverse' idea\footnote{See e.g. \cite{Greene} for a number of such scenarios.} based, for example, on the eternal  inflationary scenario \cite{Linde} or the so-called String Theory `Landscape' (see e.g. \cite{Susskind}),
are examples of such speculative UTEs.
Such scenarios are often employed in order to explain seemingly anthropic fine-tunings in some cosmological parameters 
whose values are currently not accounted for  in the so called standard cosmological model, such as the observed value of 
the cosmological constant $\Lambda$.   

An important feature of the third and to some extent the fourth of the above categories is that the number of often causally isolated (and hence {\em in principle} impossible to observe directly) UTEs they invoke is extremely large (relative to the relevant scales of the observational Universe; for example the number of elementary particles in the observable Universe), or in fact \emph{infinite}, which raises important questions about their ontological status.\footnote{It is not easy to give an uncontroversial answer to the question of which branch of physics initiated the contemporary speculative discussion of a plurality (i.e. very large or infinity) of universes. While the Everett interpretation of quantum mechanics dates back to the mid-1950s, and surely had a non-negligible influence outside quantum physics (essentially re-injecting the millennia-old idea of a plurality of worlds with scientific respectability), the concept of Multiverse which emerged after the rise, in the 1980s, of the eternal inflation scenario \cite{Linde,vilenkin}, developed in relative independence from quantum mechanical interpretations. More recently, the so called String Landscape has provided another avenue for postulating a vast number of universes, corresponding to the large number of possible false vacua allowed by String Theory. For the purposes of this paper, it will suffice to note that, being underpinned by an incomplete theory and lacking any observational support or confirmed predictions, such an idea remains speculative. For a comprehensive history of the idea of Multiverse in the XX century, see \cite{Kragh}. }
While, as realists and naturalists, we defend the theoretical indispensability and ontological reality of the first two kinds of UTEs, the aim of this paper is to ask to what extent the employment of  the latter two types of UTEs, which are postulated to be very large (in the sense made precise below), or infinite, in number are justifiable as explanatory tools in a scientific theory.

\section{Infinity in Mathematics}

Before turning to the discussion of the {\em infinite turn} in Section \ref{inf-turn},  it is worthwhile to very briefly recall the status of infinity in 
Mathematics and in conceptualisations of the physical Universe, respectively.

The intuitive idea of (countable) infinity is construed as the limit of the sequence of natural numbers $1, 2, 3, ....$ . Throughout history it has been assumed, by and large, that such infinities are asymptotic limits rather than existing limits in reality in finite time. The founding authority on this question, whose influence has extended since antiquity was Aristotle, who, in his {\em Physics}, struggled with the paradoxes posed by the concept of infinity, invoked in the context of his inquiry on time, change and division of magnitudes. In his {\em Metaphysics}, Aristotle drew an important distinction 
between {\em potential infinity} and {\em actual infinity} (i.e. between infinity as a process and infinity as a completed given cognitive object).  In line with his empiricist concerns, Aristotle argued that everything that is is infinitely divisible {\em in potentia}, and yet there is no possible infinity \emph{in actu}, which can be reached by eternal addition or subtraction (see \emph{Phys.} III.6). For Aristotle, who drew strict demarcating lines between metaphysics, physics and mathematics, the role of the physicists is to examine real entities, without applying the quantitative abstractions of the geometer and the arithmetician into their inquiry into physical reality. Aristotle's universe is a finite one, in which it is impossible to exceed every definite magnitude, for if it were possible there would be something bigger than the heavens (Phys. III.7, 207b).  

Following a rather complex history, the Aristotelian paradigm regarding infinity remained dominant for centuries, becoming refined by the late Medieval and early modern philosophers. Eventually, during the Renaissance the notion of a physically infinite cosmos was rehabilitated through the work of influential natural philosophers such as Bruno, Campanella and, particularly, Galileo whose laws of motion, developed through an anti-Aristotelian mathematised physics, challenged the finite paradigm of classical cosmology and introduced the concept of an infinite universe.\footnote{See the classic work of Alexandre Koyre, and his well-known thesis of a crucial passage, specific of modernity, from the closed world to the infinite universe \cite{Koyre}.} 

Towards the end of the 19th century, it was shown by Cantor, through his development of Set Theory, that, in addition to the countable infinity of natural numbers, it was possible to think in a mathematically rigorous way of infinities larger than the infinity of the natural numbers. An important example of such larger infinities
is the so called continuum, which can be thought of as the set of all the irrational numbers that
reside between any two whole or rational numbers. In fact Cantor showed that an unlimited hierarchy of larger and larger infinities can be constructed, the so called transfinite numbers, that grow without limit. Cantor's ideas were received with suspicion even among many mathematicians.
David Hilbert, to mention one pre-eminent example, was very critical of such numbers, as he was of the concept of infinity overall, especially in relation to the physical Universe \cite{Hilbert}. 
Our aim here is not to delve into these controversies within Mathematics, but rather to recall that even in the context of pure Mathematics there are different positions regarding the admissibility of infinity and of the Cantorian transfinite numbers. The key question for us here, however, is: what could be the status and possible relevance of such numbers for physical reality and scientific explanations? 
 
\subsection{Mathematics and Reality}

A crucial question regarding Mathematics is whether mathematical concepts, statements and theorems are discovered (i.e. exist out  there, independently of us), or are constructed (by us)? This debate has been raging in the philosophy of mathematics for decades, so far with no clear consensus. Positions on this question range from neurobiological reductionism, viewing mathematics as a cognitive product entirely contingent on the activity of our brain \cite{Changeux-Connes, Dehaene}, to 
mathematical Platonism defending the independent existence of mathematical objects, or even postulating a 'third realm' (neither physical nor mental), populated by such eternal objects \cite{Changeux-Connes, Penrose}. Others have gone even further and have directly associated physical reality with mathematical structures, arguing that 'our universe {\em is} mathematics'
\cite{Tegmark:2007ud}.

Given that the human brain is a biological construct, and hence contingently emerged from within the physical world, it would not be surprising, from a naturalist perspective, if its cognitive-neurobiological features evolved as conditioned by the structure of the Universe. For example, our observable physical Universe seems to satisfy the 2-valued logic in its macroscopic structure and dynamics. It is therefore unsurprising that we have developed a mathematics based on such a logic. To this extent, the mathematics we have is a consequence of the logic that seems to be hard wired in the structure of the physical Universe we inhabit (and, consequently, in our brains), rather than vice versa. Yet, an explanation still seems to be pending as to how mathematical concepts developed in an abstract setting can come to have, at a later time, relevance for empirical, physical, 
reality (which following \cite{Wigner}, is usually referred to as the `unreasonable effectiveness of mathematics in the natural sciences'). 

Another key question is: do {\em all} statements/concepts that could be developed within Mathematics have a counterpart in physical reality? This is a very difficult and in principle an open question, but it is certainly not at all obvious that all mathematical concepts should necessarily have a counterpart in the real Universe, particularly (and as it primarily concerns us here) in the case of a concept such as infinity, which is an asymptotic limiting concept whose application to the physical universe presents obvious conceptual obstacles (as we have learnt to acknowledge since, at least, Kant).

\section{Infinity and the physical Universe}

In discussions of infinity, and especially of its possible connection and relevance to the real Universe, the key questions to bear in mind are: (a) whether we think of infinity as an actual or potential concept, i.e. whether we have in mind an open unending process, or an already constructed infinity. Clearly the existence of the former does not necessarily require a Universe which is in some actual sense infinite at any given time; (b) The distinction between the space of potential possibilities and the space of actually existing possibilities. These are clearly not identical, as the latter is subject to the contingencies provided by the underpinning physical laws; and (c) The distinction between real infinities and extremely large numbers. In this connection it is useful to define, for a given system, including the Universe as a whole, the notion of {\em relative infinity}
as numbers that are very much larger than the size (or any other important physical characteristic) of the system, say in Planck units.

In contemporary physical theories, the concept of infinity arises in at least three distinct settings: (i) possible infinities of space and/or time, corresponding to the size (whether the extremely small or the extremely large), and the age (whether in the past or future) of the Universe; (ii) infinities arising at certain predicted limits of viable theories, such as the black holes or the cosmological singularities suggested by general relativity, and (iii) infinite ensembles used as tools for explaining observables. 

Regarding infinities of type (i), it is not clear how they can be established observationally.\footnote{This is true at least regarding the spatial extent of the Universe. It is interesting to ask if it is possible to establish observationally whether the Universe was in fact past eternal (see for e.g. \cite{mulryne-etal})} The question of future eternality does not pose a similar problem, as it will take an infinite time to arrive!\footnote{For a critique of the intelligibility of the notion of an actual, physical infinite set see \cite{StoegerEllisKirchner}. 
Note that our argument does not depend on this ontological assumption: we are more concerned with the epistemological issue of what does and what doesn't count as an explanatory hypothesis, bracketing the ontological question.} In particular, it is not clear whether it is possible,
at any point in its history, to distinguish in practice between a universe that possesses an infinite extent and one that is extremely large, say relative to its observable size.  For example, since Einstein's field equations do not constrain the overall topology of the Universe (see e.g. \cite{ellis71,levin,mota-etal}), 
the Universe could be finite spatially but appear limitless, if it has a size much larger than the present horizon.
Furthermore, even if it was possible to determine that the Universe is infinite in extent in principle (which is unlikely in practice),
it is not at all clear whether one could, in a physical context, distinguish between the
hierarchies of infinities described by Cantor's set theory, and how such infinities could be of relevance 
in a real Universe. For example, if our current understanding of the physical laws underpinning the structure and evolution of the Universe, and our attempts at formulating a theory of quantum gravity, are correct, spacetime is likely to be {\em discrete} at a fundamental level (thus not allowing infinite divisibility).  If so, even the continuum, let alone Cantor's transfinite numbers, is unlikely to be anywhere instantiated in the physical structure of the real Universe.
 
The infinities of type (ii) are generally thought to arise as a result of inadmissible applications of theories outside their domains of applicability. As a result they are often thought as temporary placeholders, to be removed (perhaps re-normalised) once appropriate corrections are made to the theory, or once a more reasonable theory of the relevant domain is formulated. On the other hand the presence of infinities of the type (iii) seem to be different in 
nature with a different methodological status, in that they are often assumed to exist by choice, in order to offer {\em explanations} for some puzzling features of the observable Universe. 

To summarise, the infinities invoked in speculative cosmological accounts of the real Universe are likely to be mostly of the type (iii) and hence voluntarily chosen (as opposed to those that arise at
less well understood  limits of classical theories, such as  general relativity, and grudgingly tolerated as temporary placeholders) and based on either the questionable trans-categorial extension 
of mathematical concepts into the physical/cosmological domain, or on metaphysical speculation, rather than motivated by empirical observations.

\section{The infinite turn}
\label{inf-turn}

Even though the idea of infinity has a relatively long history, its use in mainstream science/cosmology can be said to have accelerated and widened over the recent years. There seems to have been an intentional {\em turn to the infinite}; in the sense that attempts have been made to answer a range of fundamental questions, in both science and 
philosophy\footnote{We shall leave the discussion of infinite turn in Philosophy for a future article \cite{Gironi-Tavakol}.}, by invoking extremely large or even infinite ensembles of {\em actually} existing universes -- far (or infinitely) larger than the system to be explained. Below we give a brief discussion of some of these attempts.

\subsection{Relative or real infinities in explaining the Universe}

In traditional scientific explanations the Universe was viewed as a system with a single and connected unfolding reality.
More recently, however, it has become somewhat acceptable to explain features of our observed Universe in terms of 
(relative or real) infinite multiplicity of universes, allegedly 
covering the full spectrum of possible parameter values of interest. This development can be imputed to the encounter of two trends: 
on the one hand incomplete theories which suggest a possible proliferation of universes, and on the other 
the need to explain the seeming 
fine-tuning of certain cosmological parameters. In discussing these developments it is essential to distinguish different cases, 
depending upon whether the underlying theory employed is complete or incomplete.

\subsubsection{Many world interpretation of QM}

One of the first speculative attempts in this direction was due to Everett \cite{Everett}, who in 1957 gave an interpretation of quantum mechanics 
according to which the wave function is treated as a real and literal description of the world, thus denying its collapse in the process of observation 
to a single observed state. Instead, he postulated that all other branching outcomes are possible alternative histories, each representing an actual, 
physically real `world' (or `universe') populated by slightly different copies of observers.\footnote{For a recent philosophical defence of the Everettian hypothesis see \cite{Wallace-2012}.} This is one of the earlier examples, in physics\footnote{The idea of an infinity of possible worlds has a longer tradition in philosophical metaphysics, 
from Leibniz \cite{Leibniz} to Lewis \cite{Lewis}.}, of what might be called an {\em infinite ontological jump}, i.e. treating 
all that is potentially possible as in fact actually existing. A crucial feature of this speculative picture 
is the non-interacting nature of these `worlds', 
rendering questionable their status as entities amenable to scientific verification.
 
\subsubsection{Landscape and Multiverse}
 
The Multiverse idea has been invoked in a number of settings. In the context of String Theory,
the concept of Multiverse (see for example \cite{Susskind,Greene}) is underpinned by a so called Landscape which itself hinges 
on a number of important assumptions which is fair to say are speculative in nature, at least at present.\footnote{See, for example, \cite{ellis2011} 
for arguments by other authors not proponents of the Multiverse idea.}
These include (i) the assumption that String Theory, as it stands, 
is a complete theory of Physics and (ii) that, at least implicitly, 
there is a natural mechanism of compactification, from 10 or 11 dimensions to the observed 4-dimensional spacetime, even though such mechanisms are not known at present, within the theory itself.
Regarding (i), it is fair to say that String Theory, despite its 
successes in the direction of unifying Quantum Physics and Gravity,
is not a completed theory with clearly verified observable predictions. 
Clearly until it is established as a completed theory, with clear testable predictions, taking seriously its speculative potential consequences is problematic (see Sec \ref{closure} below for 
further discussion). Concerning (ii), as far as we are aware all attempts at finding natural and generic mechanisms for compactification within String Theory 
itself have so far been unsuccessful. This absence, were it to persist, is potentially very problematic as it might indicate that such a
mechanism may require a meta-theory that would transcend String Theory itself, as it is currently understood.
Furthermore, there is no guarantee that such a mechanism(s) would result in the vast number of viable 
compactifications, which are similar to our Universe, as their end states\footnote{It is known that dynamical systems can have multiple attractors, but it is extremely difficult to imagine 
a dynamics arising from a physical theory which gives rise to $10^{100}$ to $10^{500}$ number of compactified end states. In fact it is very 
likely that the number of viable compactifications is in fact far smaller, which raises the question of how they are selected!} 
which is what is normally assumed, at least implicitly, in justifying the Multiverse scenario. In any event even if such enormous {\em multi-attractor} 
settings exist, it is unlikely they could all be realised simultaneously. In classical settings, dynamical systems 
can possess multiple attractors. In that case, however, which attractor the system tends to will depend 
on which basin of attraction the initial conditions are set in. Thus in the Landscape setting the problem of why this compactification and not another 
could be related to how the initial conditions are set. In the quantum settings the problem is more complicated and the outcome depends on how the 
probability decay rates are defined, which again may point to an external meta theory/mechanism. 
 
 The other setting in which the concept of Multiverse is invoked is in the context of so called eternal inflation. Again even though inflation has had great success, particularly in explaining structure formation in the early Universe, it is fair to say that (rather than being a theory) it remains a (set of) model(s) which as yet have not been successfully embedded within a complete theory of fundamental interactions.

To summarise, the Multiverse scenario is an example of an idea based on potential speculative outcomes of an incomplete theory under construction and without unambiguous empirical confirmation (we question whether the consequence of an incomplete theory can be legitimately called a `prediction')\footnote{Compare the way in which inflation `predicts' eternal inflation and the formation of an infinite number of universes with the way in which General Relativity predicted the (now observationally confirmed) existence of gravitational waves, or the way in which the wave theory of light predicted a bright spot at the centre of a shadow cast by a round object. Note that inflation itself was originally postulated as an inference to the best explanation, in order to account for the observed lack of magnetic monopoles, the flatness of spacetime, and the homogeneity and isotropy of the observable universe. It aims at doing so by offering the description of a physical mechanism responsible for inflation. While 1) the fundamental theory underpinning such a  mechanism is not yet fully understood, and 2) there remains the possibility that other non-inflationary explanations might be given in the future for the observations that originally motivated it, we look favourably upon it because, unlike multiverse theories, inflation offers a speculative {\em finite causal explanation} (see the next section).} of where the theory is implicitly taken to be complete and its speculative possible consequences taken as ontologically real.

\section{Closure, Empirical Testing and Explanation}
\label{closure}

The notion of {\em closure} plays a pivotal role in various employments of the 
notion of infinity in scientific and philosophical explanations.
In the case of the former, such as in Many Worlds, Anthropic or Multiverse reasonings,
there is an urge to implicitly assume that our present understanding of the laws of nature (and the Universe) is 
complete, in what amounts to an assumption of `closure'\footnote{This urge at having  a `total' explanation
of reality, which is implicitly assumed to be complete and therefore closed,
has had a long history in human history.} (For a fuller discussion of the question of closure in physical theories see \cite{Tavakol-Anderson-2015}), which we define as the assumption that laws of Physics as they are currently known are complete, irrespective of future observations.
Interestingly, in debates between proponents and critics of the idea of a Multiverse, the accusation of scientific 'conservatism' is often directed by the former group against the latter, indicating an alleged unwillingness to entertain new speculative ideas and to extend our idea of what counts as legitimate sciences, coupled with an excessive reliance on present-best scientific consensus, bordering on in-group thinking. It is questionable wether stricter adherence to the epistemic constraints of theory evaluation (most of all, that of empirical testability and predictive success) can count as 'conservatism', since only criteria of correctness warrant the possibility of detecting an inconsistency in one's theoretical approach.\footnote{A recent  intervention on the topic of scientific method is that of Dawid \cite{Dawid}, who in the specific context of a defence of the scientific respectability of String Theory,  makes the case, more generally, for an enlargement of our criteria for the acceptance of a theory to encompass non-empirical ones. Briefly, Dawid proposes three standards for non-empirical assessment: (1) the No Alternatives Argument, holding that our acceptance of the adequacy of a new theory should increase where no plausible alternatives are present; (2) the Unexpected Explanations Argument, holding that our acceptance of the adequacy of a new theory should increase where it is able to offer explanations for phenomena it was not originally formulated to deliver; (3) the Meta-Inductive Argument, holding the more a new theory is capable of being embedded into a whole of currently accepted and established theories the more we are entitled to believe it. We leave a more detailed discussion of this proposal for a later publication, as it seems to us that a key problem with this it is that, in the absence of experimental verification, there is a danger that the determination of whether these criteria are {\em satisfied} becomes subjective. However, provisionally accepting these criteria for the sake of argument, it seems to us that  {\em even then} the multiverse hypothesis fails, not meeting criteria 1, 2 and arguably even 3.}

  Progress is achieved via the correction of mistakes, requiring well-defined criteria for spotting such mistakes. There is a slippery slope going from speculation-friendly methodological liberalism to a Feyerabendian `anything goes' approach. The doubt can also be raised that, perhaps, it is rather these speculative proposals that can be termed `conservative'  by implicitly assuming  closure -- while in practice employing an incomplete theory. We are not against speculation in science (and philosophy), indeed we believe it to be an indispensable ingredient for the development of constantly revised and increasingly accurate conceptual frameworks.  
However, the character of a speculative proposal should be that of opening up inquiry by assuming a thoroughgoing fallibilism vis-a-vis current orthodoxy and 
acknowledging protocols for the acceptance of new proposals.

Indeed, It seems to us that, today, all too often reference to infinity in cosmological theorising is motivated 
by a hasty urge to treat our present theories as complete (at least implicitly)
and is thus employed as an {\em unexplained explainer} to make sense of (well-known) puzzling observations. If it is the case that, 
broadly speaking, a scientific explanation is one that aims at reducing the number of `brute',  or unexplained facts, to compress the 
information needed to offer a description of the physical process, it seems to us that to do so by invoking a single major `brute fact' 
such as the Multiverse, fails to achieve the desired descriptive economy. The aim of a cosmological explanation should be that of 
providing a deeper understanding of the Universe, but the Multiverse type scenario rather shifts the target of the explanatory task 
from the finite observed Universe  to a postulated relative or real infinite `Landscape'. Indeed, while we endorse a pluralistic approach 
to the concept of explanation, mindful of the difference it displays across scientific disciplines, it appears that the `Multiverse explanation' 
is an entirely  {\em sui generis} kind of explanation, failing to offer either a nomological (it is debatable wether the Multiverse scenario really is a consequence of a complete theory),
 or a causal (Multiverses would be causally disconnected, and the physical process responsible for their creation is, again, at best speculative) kind of explanation for the observed Universe. We are not against the careful use of (`weak') anthropic arguments in scientific explanation, as kind of `cosmological transcendental arguments' but we caution against over-extending their use: if observed variable X is a condition of possibility for the very existence of the observer, it is trivially true that the existence of a life-allowing universe in general, and of the observed value of X in particular, is `explained' by anthropic constraints. The problem is that the {\em causal processes} (spatiotemporally contiguous processes capable of transmitting information) \cite{Salmon1, Salmon2} which led to the existence of the observed value remain unexplained, and indeed the {\em explananda} (these observed values) are not put in any relation of causal dependence (no matter how indirect) with the {\em infinite} ensemble of  universes. Anthropic considerations `explain' why we exist (and do science) in a life-friendly universe, but they do not explain why and {\em how} do life-friendly universes exist. So, in the case of the Multiverse scenario, a causal explanation outlining a clear chain of causal relations is replaced by an anthropic probabilistic argument which rests on the problematic assumption of an extremely large (or an actual infinity) of combinations of physical laws, fundamental parameters and initial conditions. This is at best a {\em speculative description} of the universe we live in and its history (answering the question `what is') but hardly an {\em explanation} (answering the question `why?' by tracking of the causal structure of the physical universe).\footnote{We here subscribe to a notion of explanation as essentially tied to causation. Wesley Salmon is the philosopher traditionally indicated as re-injecting the concern with causation in the philosophy of explanation, under the general principle that `explanatory knowledge is knowledge of the causal mechanisms...that produce the phenomena with which we are concerned' \cite{Salmon1}. A more recent account, arguing for a manipulatist account was defended by Woodward \cite{woodward2003}, according to whom the distinction between description and explanation resides in the latter offering information in principle useful for manipulation. While the practicality of manipulation is obviously out of the question in a cosmological setting, Woodard observes that the notion can be generalised through the notion of an impersonal `intervention' and understood counterfactually: `an explanation ought to be such that it can be used to answer what we call a {\em what-if-things-had-been-different} question: the explanation must enable us to see what sort of difference it would have made for the {\em explanandum} if the factors cited in the {\em explanans} had been different in various possible ways, that is to say if an intervention in cause X would have brought about a change in effect Y. The Multiverse scenario clearly fails to meet this criterion for being an explanation: the conditions leading to the present universe in fact were (are?) distributed across a very large (or infinite) space of actually existing possibilities where either everything vacuously explains everything else, or nothing explains anything at all.}

The unwarranted assumption of closure which seems to motivate models with physical infinities immediately poses another important question: how is it to be reconciled with the openness displayed by the {\em history} of our cosmological theories? Our understanding of the universe has changed dramatically over the last few centuries, and while our current understanding of it is demonstrably more profound and accurate than it was just half a century ago, there is no reason to believe that the unexplained features and parameters in the current `standard model' of cosmology, which are taken to justify the use of physical infinities, will not be explained by a future physical (cosmological) theory without the need for such infinities.
 
An example of argument proposed in support of the Multiverse proposal is the `historical-progressive' argument \cite{Carr2014}, according to  which the Multiverse scenario is a natural, or even necessary, step along the path of de-centring of humanity from a privileged and central position, leading from the Copernican revolution to a contingently assigned placement for humanity, and our observable universe, along an infinite spectrum of universes. While prima facie appealing, we find this type of argument unconvincing.
Of course we strongly endorse the humbling effect that cosmological inquiry has had (and should keep having) on our self-understanding.\footnote{Thus we want to ward ourselves against a potential 
misreading of our argument: we fully endorse the secularising drive of cosmological de-centring, and the ejection of anthropocentric notions from physical cosmology, but these cannot be used as a {\em premise} for scientific inquiry).} Nevertheless, we doubt that there is anything more than a passing resemblance between the post-Hubble abandonment of a 'galactocentric worldview' and that of a 'cosmocentric' one (in favour of a Multiverse scenario): the two theoretical shifts are warranted by quite different methodological principles. The revolutions referred  to in this meta-historical line of argument originated as explanatory conjectures but were subsequently unambiguously empirically verified, and remain today object of continuous observational confirmation. Consider, for example, how the heliocentric picture was slowly reached inductively, by painstakingly tracking the behaviour of the celestial bodies, ultimately preferred to the Ptolemaic theory according to criteria of simplicity and empirical adequacy, and much later also `observationally' confirmed when the first imaging probes were sent in orbit around the Earth. Or recall how the 'great debate' between Shapley and Curtis \cite{Hoskin, Kragh} regarding the location of the spiral 'island universes' was finally, unambiguously, and observationally settled by Hubble's distance measurements based on the luminosity-period measurements of the Cepheid variables: a conclusion supported by well-understood physical laws and causal mechanisms.
The key distinguishing feature of the Multiverse type proposals is that there is no possible observational evidence capable of confirming, or importantly ruling out, a (finite or infinite) Multiverse, and hence of putting an end to this analogous twenty-first century 'great debate'. To endorse the Multiverse scenario and postulate an extremely large or an infinity of causally disconnected universes in order to retroactively explain puzzling data is not a warranted extra step to make for those committed to the epistemic priority of physical cosmology for the description of our place in the cosmos. It unquestionably involves a relaxation (or, more charitably, a revision) of our epistemic standards and of accepted criteria for the scientificity of an hypothesis. Abductive explanations, inferences to the best explanation, are valid forms of scientific reasoning (generating plausible hypotheses from incomplete information) but only insofar as (a) they are able to offer a causal story (a possible explanation is a possible causal story) and (b) the proposed explanation can be defensible in light of new, more or less direct evidence (suggesting another, better causal story). The Multiverse hypothesis seems to fail on both counts. To reiterate: we are not merely arguing that Multiverse speculations go beyond science because of methodological shortcomings (as this has been the main avenue of critique so far), but pointing out how they do not seem to offer any kind of {\em explanation} that would count as delivering {\em scientific understanding} (an understanding with both a theoretical and a practical component): the goal, we presume, of any scientific endeavour. 


\section{Economy of explanation, information and infinity}

In assessing scientific explanations of finite real systems an important question 
concerns their efficiency.  This is in principle rather a difficult concept to quantify, but as a potential first step a possible criterion that may be 
considered is the  `relative information content' of the explanation. There are many ways one may try to do this. One possible way could be to proceed
in analogy with Kolmogorov's notion of complexity, according to which the complexity of a sequence, $S$,
corresponds to the length of a program needed to produce that sequence. Thus a sequence
is called random if the length of the shortest program required to encode it is the same as the length of the sequence itself.
Now consider a system $U$ and let
the information content of the system be $I_u$. Now consider an explanation of this system whose information content is $I_e$.
One could say the explanation is {\em reasonable} if $I_e \sim {\mathcal{O}} (I_u)$.
A crucial question in this connection is how much larger than $I_u$ can $I_e$ be 
and still be treated as a reasonable explanation? 

To fix ideas, consider our observable Universe as an example.  According to our present understanding in cosmology, the number of elementary particles in the
observable Universe is $\sim \mathcal{O}$$(10^{80} )$. 
Bearing this in mind, an important question is what would be the physical
status of (relatively infinite) numbers such as $10^{100}$ or $10^{500}$ in such a  universe?
 
The question then is, would an explanation of the current Universe, with its  $\sim \mathcal{O}$$ \left (10^{80}\right )$ particles
 make sense if it involves ensembles or numbers of constituents which are relatively infinite (say $\sim \mathcal{O} $$(10^{100})$ or $(10^{500})$) or in fact infinite in size?

Of course the existing Universe may in fact be infinite. The question then becomes whether such an infinity 
can ever be operationally decidable/determinable in finite time,
and how to associate information with such a universe.

To summarise, for an explanation of a system to be reasonable, one would expect the information content of the explanation not to 
enormously (or infinitely) exceed the information content of the system under consideration. In that case it would be doubtful 
whether the concept of (relative or real) infinity, or worse still that of Cantor's transfinite numbers are adequate as meaningful 
tools to be employed in order to scientifically explain finite aspects of, or the totality of, the real (finite) observable Universe.

\section{Why infinity may not help after all}

In the majority of instances in which the concept of infinity (such as an infinite set
of values for a parameter or an infinite ensemble) is invoked as a tool for explaining some aspect of the Universe,
it is implied that the assumed infinite set necessarily includes 
the space of {\em required possibilities},
including occurrences as close to or indistinguishable
from those required to explain the feature under consideration.

A key point often ignored in such attempts is that the {\em required
infinity} is not just any infinite set! Take for example a parameter which may be required 
to explain the current observed state of the Universe, such as 
for example the required value of the
cosmological constant\footnote{A non-zero cosmological constant of an appropriate
size could in principle account for the observed puzzling late acceleration of the Universe,
while being compatible with all other known cosmological observations. 
A crucial question is why does this constant take the value it has.
Here we are leaving aside the fundamental
question of whether this is the only way to explain the late acceleration of the Universe 
(see e.g. \cite{clifton-etal1, tsujikawa,clifton-etal2} for 
other scenarios for possible explanations of the late acceleration).}, $\Lambda$.
Imagine the real line as the axis along which the cosmological constant
can take its values. Now any finite neighbourhood
of the real line contains an infinite (continuum) set of values. So having infinite
possible values for the cosmological constant is not sufficient for it 
to take its observationally required value. We still need an infinite amount of fine tuning
in order to have the infinite set of possible values of the constant to lie 
in the required neighbourhood of the real line.

\section{Conclusion}

We have considered the employment of concept of infinity in various scenarios employed over the recent decades to 
explain a number of questions in science, particularly in cosmology.
By introducing the notion of relative infinity and the idea of closure, and employing the 
concept of information content of an explanation, we have argued that scientific explanations of the
finite observable Universe (or any finite system) involving 
infinities (of ensembles or Universes) are likely to be problematic, or in fact fail to be {\em reasonable explanations}. While speculative ideas and creative leaps have played an undeniably crucial role throughout the history of science (and are likely to do so in the future), scientific theorising should follow some criteria of empirical meaningfulness, and its proposed explanations need to follow an agreed-upon template of what counts as an explanation in a given theoretical domain.\footnote{We are not claiming that {\em all} explanations need be causal explanations. But we are claiming that, in fields like fundamental physics and cosmology causal explanations in fact {\em are} the kinds of explanation which are required for an understanding of the phenomena.}  We do not reject the Multiverse scenario because of its speculative nature, but because it fails to be a genuinely explanatory hypothesis and to predict any new phenomenon -- in other words, failing to contribute to our scientific {\em understanding} of the universe. The replacement of a benevolent, parameters-tuning deity with an unobservable, and causally isolated infinite (transfinite?) set of universes is, surely, a step in a naturalistic direction, but one trading faith in the supernatural for a `natural' explanation which requires a state of affairs enormously more complex than the observed Universe. Neither choice seems satisfactory. Science proceeds via trial and error, observation and theory-construction, and in this context speculative hypothesis should be considered when they offer an alternative {\em finite causal mechanism} which can lead to the explanation of the phenomena under consideration.
 
\section*{Acknowledgments}

We would like to thank Andreas Brandhuber, Peter Cameron, Bernard Carr, George Ellis, Roy Maartens and Sanjaye Ramgoolam for helpful discussions and comments.




\begin{thebibliography}{}
\bibitem{Osnaghi-etal-2009} S Osnaghi, F Freitas, O Freire Jr, `The origin of the Everettian heresy'
Studies in History and Philosophy of Modern Physics (2009)


\bibitem{Everett} H.~Everett, in `The Many-Worlds Interpretation of Quantum Mechanics', B. S. DeWitt \& N. Graham (eds.), Princeton Univ. Press, Princeton (1973)

\bibitem{Greene} B.~Greene, `The Hidden Reality: Parallel Universes and the Deep Laws of the Cosmos', Penguin Books (2011)

\bibitem{Linde} A.~D.~Linde, `Eternally Existing Self-Reproducing Chaotic Inflationary Universe', Phys. Letts. B 175, 395 (1986)

\bibitem{Susskind} L.~Susskind, `The anthropic landscape of string theory', arXiv:hep-th/0302219

\bibitem{vilenkin} A.~Vilenkin, `Birth of Inflationary Universes',  Phys. Rev. D 27,  2848 (1983)

\bibitem{Kragh} H.~Kragh, `Higher Speculations: Grand Theories and Failed Revolutions in Physics and Cosmology', Oxford University Press (2011)

\bibitem{Koyre} A.~Koyre, `From the Closed World to the Infinite Universe', Baltimore: Johns Hopkins Press (1957)

\bibitem{Hilbert} D.~ Hilbert, `On the Infinite'. In Philosophy of Mathematics. Ed. P Benacerraf and H Putnam (Englewood Cliff, N. J.: Prentice Hall), pp 134-15,  (1964)

\bibitem{Changeux-Connes} J.~P.~Changeux, A.~Connes, `Conversations on Mind, Matter, and Mathematics', Princeton University Press (1995)

\bibitem{Dehaene} S.~Dehaene, `The Number Sense: How the Mind Creates Mathematics', Oxford University Press (2011)

\bibitem{Penrose} R.~Penrose, `The Road to Reality: A Complete Guide to the Laws of the Universe', Johnathan Cape (2004)

\bibitem{Tegmark:2007ud} M.~Tegmark, `The Mathematical Universe',' Found.\ Phys.\   38, 101 (2008)

\bibitem{Wigner} E.~Wigner, `The Unreasonable Effectiveness of Mathematics in the Natural Sciences', Communications in Pure and Applied Mathematics, XIII (1960)

\bibitem{mulryne-etal} D.~ J.~ Mulryne, R.~ Tavakol, J.~ E. Lidsey, G.~ F.~ R.~ Ellis, `An emergent universe from a loop', Phys. Rev. D71, 123512 (2005)


\bibitem{StoegerEllisKirchner} W.~R.~Stoeger, G.~F.~R.~Ellis and U.~Kirchner, `Multiverses and cosmology: Philosophical issues' (2004) [astro-ph/0407329v2]

\bibitem{ellis71} G.~ Ellis : Gen. Rel. Grav., 2 , 7 (1971)

\bibitem{levin} J.~ Levin, `Topology and the Cosmic Microwave Background', Phys. Rept., 365, 251 (2002)

\bibitem{mota-etal} B. Mota, M.J. Reboucas, R. Tavakol, `Circles-in-the-sky searches and observable cosmic topology in the inflationary limit', Phys. Rev. D78, 083521 (2008)

\bibitem{Gironi-Tavakol} F.~Gironi, R.~Tavakol, `Infinite turn in philosophy', in progress

\bibitem{Wallace-2012} D.~Wallace, `The Emergent Multiverse: Quantum Theory According to the Everett Interpretation', Oxford University Press (2012)

\bibitem{Leibniz} G.W.~Leibniz, `Philosophical Essays', Hackett (1989)

\bibitem{Lewis} D.~Lewis, `On the Plurality of Worlds', Basil Blackwell (1986)

\bibitem{ellis2011} G.~Ellis, `Does the Multiverse Really Exist?', Scientific American, 305 (2), 38 (2011)

\bibitem{Tavakol-Anderson-2015} R.~Tavakol and E.~Anderson, Submitted for publication (2015) [arXiv:1506.07928 [gr-qc]]

\bibitem{Dawid} R.~Dawid, `String Theory and the Scientific Method', Cambridge University Press (2013)

\bibitem{DawidRes} R.~Dawid, `'Simple' or `Elegant' Criteria are not Valid', Nature, 518 (2015)

\bibitem{Salmon1} W.~Salmon, `Four Decades of Scientific Explanation', University of Minnesota Press (1989)

\bibitem{Salmon2} W.~Salmon, `Causality and Explanation', Oxford University Press (1998)

\bibitem{woodward2003} J.~Woodward, `Making Things Happen: A Theory of Causal Explanation', Oxford University Press (2003)

\bibitem{Carr2014} B.~ Carr, In `Mathematical Structures of the Universe', ed. N.Eckstein, D. Heller, S. Szybka, Copernicus University Press (2014)

\bibitem{Hoskin} M.~A.~Hoskin, `The `Great Debate': What Really Happened', Journal for the History of Astronomy, 7 (1976)

\bibitem{clifton-etal1} T.~ Clifton,  P.~ G.~ Ferreira, A.~ Padilla, C.~ Skordis, `Modified Gravity and Cosmology, Phys. Rep. 513,1 (2012)

\bibitem{tsujikawa} S.~ Tsujikawa, `Modified gravity models of dark energy',  Lect. Notes Phys., 800, 99 (2010)

\bibitem{clifton-etal2} T.~ Clifton, K.~ Rosquist, R.~Tavakol, `An exact quantification of backreaction in relativistic cosmology', Phys. Rev. D 86, 043506 (2012)


\end{thebibliography}
\end{document}